# Far-Field Plasmonic Resonance Enhanced Nano-Particle Image Velocimetry within a Micro Channel

Zhili Zhang*, Quanshui Li, Sara S. Haque, Mingjun Zhang

Department of Mechanical, Aerospace and Biomedical Engineering, University of Tennessee, Knoxville TN 37996

*To whom correspondence should be addressed: Zhili Zhang (zzhang24@utk.edu)

## Abstract

In this paper, a novel far-field plasmonic resonance enhanced nanoparticle-seeded Particle Image Velocimetry (nPIV) has been demonstrated to measure the velocity profile in a micro channel. Chemically synthesized silver nanoparticles have been used to seed the flow in the micro channel. By using Discrete Dipole Approximation (DDA), plasmonic resonance enhanced light scattering has been calculated for spherical silver nanoparticles with diameters ranging from 15nm to 200nm. Optimum scattering wavelength is specified for the nanoparticles in two media: water and air. The diffraction-limited plasmonic resonance enhanced images of silver nanoparticles at different diameters have been recorded and analyzed. By using standard PIV techniques, the velocity profile within the micro channel has been determined from the images.

## 1. Background

In the area of fluid mechanics and aerodynamics, velocity field is one of the most important physical parameters [1, 2]. However the velocity measurement is not trivial, especially for boundary layers or micro channels in which inherent compressibility effects, temperature and density gradient make the direct measurement of velocity somewhat more difficult. The existing flow diagnostic techniques are not well suited for such measurements. Pilot-static pressure probes, Hot Wire Anemometry (HWA) [3], Laser Doppler Velocimetry (LDV) [4, 5], Molecular Tagging Velocimetry (MTV) [6-9], Particle Image Velocimetry (PIV) [2, 10-12] and micron resolution PIV (μPIV) [13-16] etc. have many limitations for the purpose of such measurement. The pilot-static probe is ineffective because of the inherent wave structure created by the probe. It also requires additional measurements of temperature or density to calibrate the flow conditions. The hot-wire probe becomes sensitive to the local mass-flux and total temperature. Hence, data acquisition and data-reduction procedures are subsequently more challenging, and velocity cannot be measured directly. Although non-intrusive laser-based techniques, such as LDV, PIV and μPIV can provide direct measurement of the velocity in such flows, all methods require seeding particles of large dimensions added to the flow. Due to the inertial effects of the particles, they may not be able to track the flow accurately, especially across the shock wave or the near wall region within the boundary layers. PIV has a significant degrading effect with increasing particle size [17]. As a comparison, MTV tags air or other molecules themselves and thus is able to follow rapid transitions: a shock wave in the flow. Of course no global information can be obtained through MTV due to the limited tagging regions. In a universal point of view, seeding particles in MTV are molecules which are on the order of a few Angstroms and have perfect tracking property. While particles used in LDV and PIV are artificially injected which are on the order of 100nm to tens of microns, they may not be able to track the flow perfectly.

A unique opportunity to resolve some of these difficulties is offered by a novel experimental technique, nano-Particle Image Velocimetry (nPIV). Nanoparticles have many unique properties: some of them may not even be well described by traditional theories in physics and mechanics due to the breakdown of the scale law [18]. Among the properties, the plasmonics-enhanced light scattering and fluorescence emission can significantly improve nanoparticles' visibility without increasing their size and weight [19, 20]. According to Stokes' theorem, as particles dimension and weight decrease, particle response time to the flow conditions becomes shorter and flow tracking capability gets better [21]. For example, PIV measurement have been successfully conducted in hypersonic flows (M = 7) by $TiO_2$ nanoparticles at diameters of ~50nm [22] and shock boundary layer interactions by $TiO_2$ nanoparticles at diameters of ~170nm [23], which are not possible at all for common micron-sized particles. However, the spatial resolution of those measurements is still insufficient to determine the near wall flow properties. Smaller nanoparticles will be able to track the boundary layer flows at a more detailed manner. As an example, natural seeding of condensed $H_2O$ and $CO_2$ particles at diameters of ~10 nm has been successfully used in the hypersonic boundary layer measurement [24]. Obviously natural seeding is limited to flows at certain conditions and cannot approach the near wall region of the supersonic boundary layers. This is because the condensed $H_2O$ and $CO_2$ particles are evaporated in the region at high temperatures.

## 2. Flow Seeders

### 2.1 Size of the seeders

First, light scattering from a particle, either Mie or Rayleigh scattering, can be calculated from Mie scattering theory, which reveals that light scattered by a particle is proportional to the square of its volume. The cross section σ of light scattering can be expressed as [25, 26]

$$\sigma = 24\pi^3 V^2 \left| \frac{\varepsilon - \varepsilon_m}{\varepsilon + 2\varepsilon_m} \right|^2 \tag{1}$$

where V is the volume of the particle, $\varepsilon$ is the dielectric constant of the particle, and $\varepsilon_m$ is the dielectric constant of the surrounding. The light scattering power $P$ is the product of the cross section and incident light intensity $I$, $P = \sigma I$. So in the light scattering, large particles always dominate smaller ones. When radius of the particles decreases by a factor of 100, for example particles at diameters of 10 nm instead of 1 μm, the scattering power will decrease by $10^{12}$. The light scattering from the nanoparticles is weak and difficult to be measured. It will be extremely hard to collect light scattering from nanoparticles if large dust particles are present. Table 1 shows typical scattering cross sections of particles as a function of the particle size. *In general, the smaller the particle is, the more difficult for it to be detected by Mie/Rayleigh scattering.*

Table 1. Scattering cross sections of particles as a function of diameters

| Particles diameters | Scattering cross section (cm$^2$) |
| --- | --- |
| Air molecules at 0.3nm [26] | $10^{-26}$-$10^{-28}$ |
| $CO_2$ cluster at 10nm [27] | $10^{-17}$ |
| Seeding particles at 1 μm [21] | $10^{-8}$ |
| Seeding particles at 10 μm [21] | $10^{-5}$ |

Second, tracking capability of seeding particles can be estimated by Stokes' theorem. The basic assumptions of Stokes' flow are that flow is incompressible and Reynolds number is very small. Thus the inertia terms can be neglected in the Navier-Stokes equations, and only the viscous force balances the pressure that drags the particle. The supersonic flow is compressible and temperature gradient in the boundary layer causes variations in viscosity. However, the local flow velocity fluctuation relative to the moving particle is still subsonic. Local Mach number fluctuation from its mean value is still below 0.3 when the free stream Mach number reaches 4 and below 1 when the Mach number reaches 7.2 for a zero-pressure-gradient adiabatic boundary layer at moderately high Reynolds numbers [28]. So Stokes' theorem can still be used to approximate and analyze how the nanoparticles follow the supersonic flows. The relaxation time of the particle can be expressed as [21]

$$\tau_p = d_p^2 \frac{\rho_p}{18\mu_f}(1+2.7Kn_d) \qquad (2)$$

Where $d_p$ and $\rho_p$ are the diameter and density of the particle, respectively, $\mu_f$ is the viscosity, $Kn_d$ is the Knudsen number. Approximately the response time of the particle is proportional to the density of the particle and the square of the diameter of the particle. Air molecules tagged by MTV can perfectly measure the velocity across the shock, while particles seeded in PIV show degraded effects. *Therefore the smaller and lighter the particle is, the better for it to track the flow motion.* The lower limit for the seeding particle is set by the measurement uncertainties due to Brownian motion of the particles. [14, 15]

## 2.2 Plasmonic Resonance Enhanced Light Scattering

Plasmonic effect is the generation of highly localized light fields in the near-field of metallic nanostructures [29]. Light scattering, including Rayleigh and Raman scattering, can be significantly enhanced by the plasmonic resonance. To quantify the plasmonic resonance effects, Discrete Dipole Approximation (DDA) method [30] has been used to calculate the light absorption and scattering from various nanoparticles in solution and in air.

In Figure 1, comparison between silver nanoparticles with TiO$_2$ nanoparticles at the same sizes is shown. With a diameter of 22.3 nm, silver nanoparticles show the resonance enhanced light scattering at 360nm in air and ~400nm in water. TiO$_2$ nanoparticles show no resonance effects.

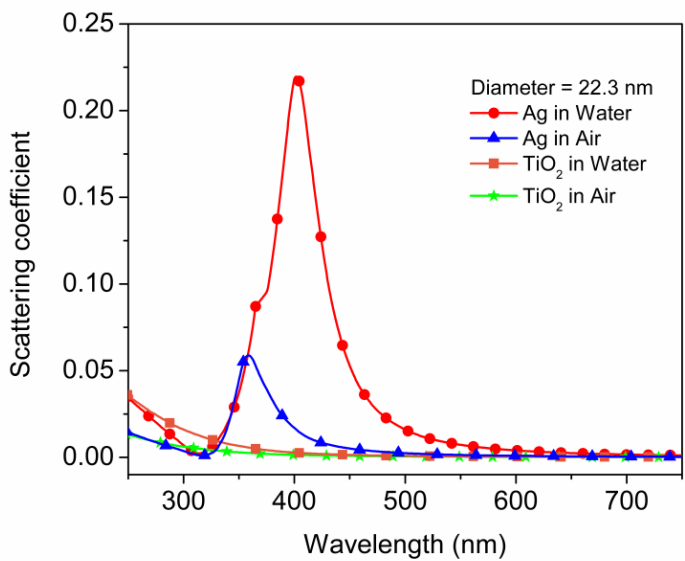

*Figure 1. Comparison of light scattering spectra between silver and TiO$_2$ nanoparticles at diameters of 20nm in air and in water*

In Figure 2, comparison between experimental and computational results of nanoparticle scattering is shown. In Figure 3, comparison of light scattering spectra among silver nanoparticles at different diameters in water is shown. The resonance peaks shift toward the red region and get broader when the diameters of the nanoparticles become larger. The position of the peaks and the width of the peak can be used to estimate the diameters of the nanoparticles. Light scattering intensity can be plasmonic resonance enhanced if the incident light is around the resonance frequency. The cross sections for the silver nanoparticles can be resonance enhanced by more than 10 times when the incident light varies from ~300nm to ~400nm in water. Since it is only possible for the silver nanoparticles, light scattering from other particles can be filtered or blocked efficiently. This will eliminate other potential disturbances from the dust particles for the velocity measurement.

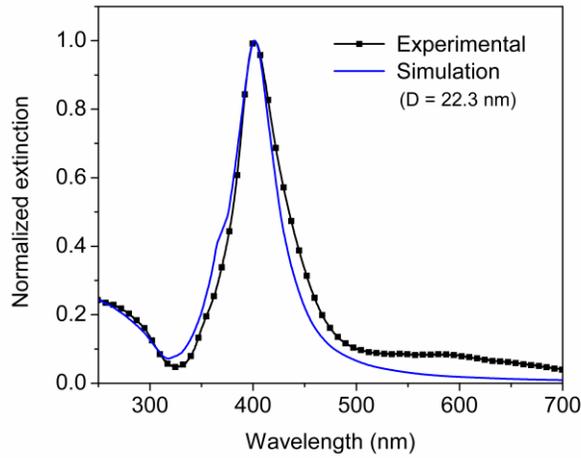

*Figure 2. Comparison of experimental and simulation results of silver nanoparticles.*

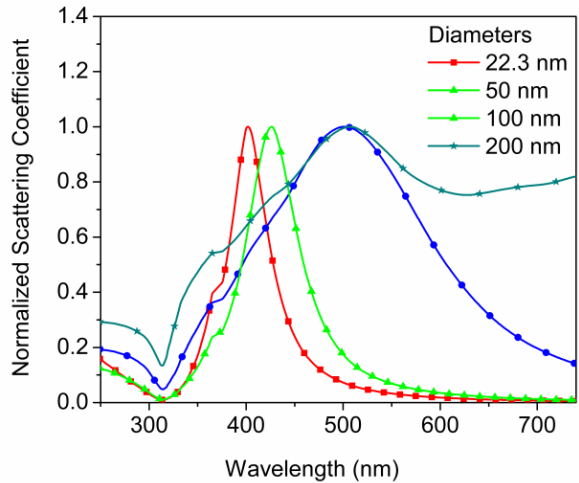

*Figure 3. Comparison of light scattering spectra among silver nanoparticles with different diameters in water.*

# 3. Nanoparticle Synthesis and nPIV

## 3.1 Nanoparticle Synthesis & Characterization

Silver nanoparticles are synthesized by chemical reduction method. First, $AgNO_3$ (90 mg) is dissolved in 500 mL of deionized water. After the solution is boiling, 10 ml of 1% sodium citrate (0.22 g of trisodium citrate, $C_6H_5Na_3O_7$, dissolved in 22 mL of $H_2O$) is added into it. The color of solution gradually turns gray in a few minutes. The mixture solution is kept boiling for 1 hour. Then the production containing silver nanoparticles is irradiated by an intensive laser beam (532 nm, 393 $mJ/cm^2$) for about 30 minutes. The silver nanoparticles are melted by the laser beam to form the nanoparticles at about 8-10 nanometers. These nanoparticles are used as seeds for further synthesis. After the same amount and concentration of $AgNO_3$ solution is boiled again, 8.3 ml of seeds solution and 10 ml of 1% sodium citrate are added into them. Keeping boiling for 1 hour, the mixture solution is cooled to the room temperature. The Ag nanoparticles in the solution are finally filtered with 100 nm filters and used for the nano-PIV measurement.

Hitachi S 3500 SEM (Scanning Electron Microscope) is used to characterize the actual morphologies of silver nanoparticles. The nanoparticles have spherical shapes. The diameters of those nanoparticles are approximately 35 nm. The silver nanoparticles used in the experiments are measured, shown in Figure 4. The size distribution of nanoparticles is counted, shown in Figure 5.

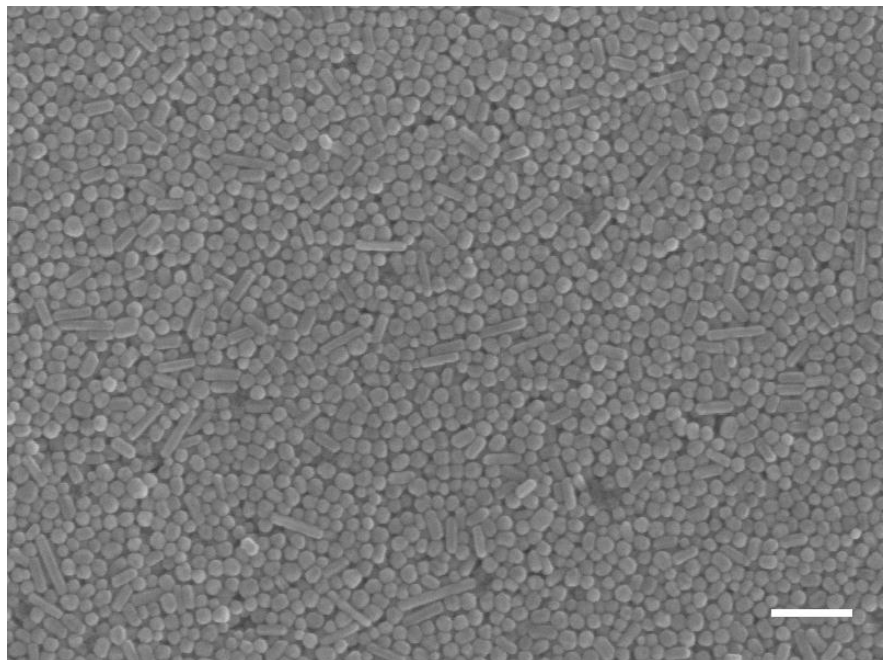

*Figure 4. Morphologies of dry silver nanoparticles record by SEM. The bar represents 100nm.*

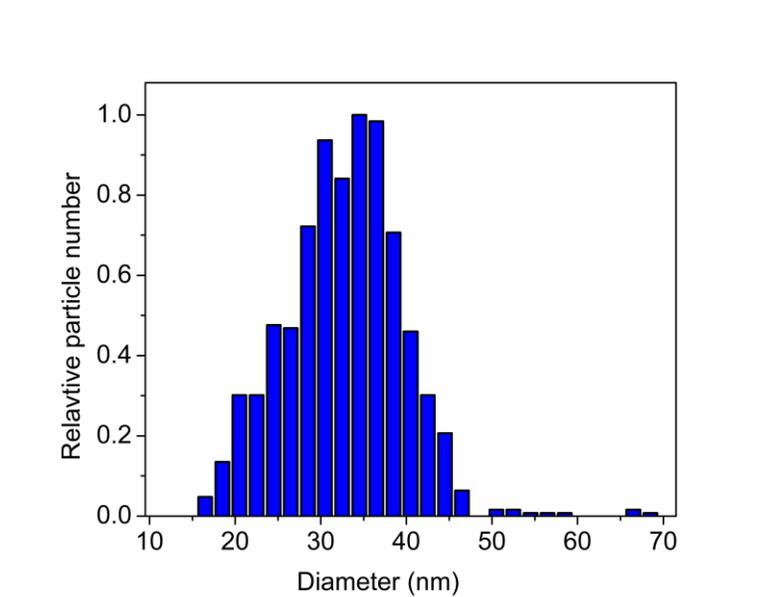

*Figure 5. The size distribution of silver nanoparticles measured by SEM*

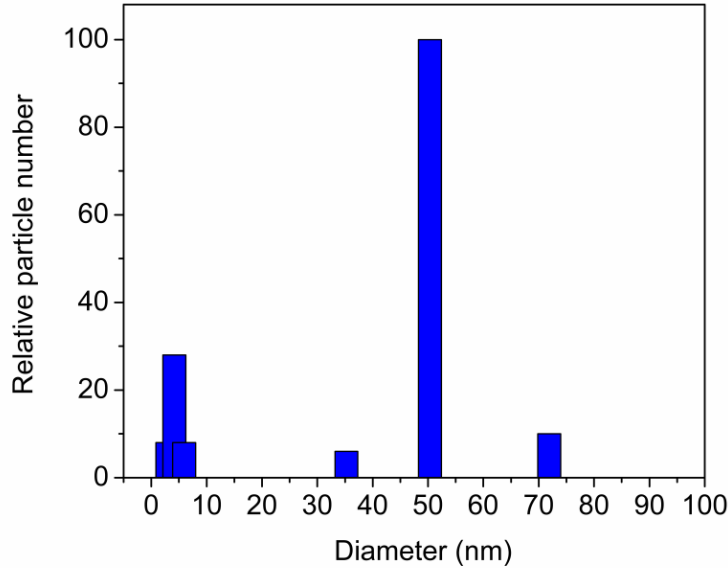

*Figure 6. Size distribution of silver nanoparticles in solution measured by DLS*

Size distribution of the nanoparticles is recorded by a dynamic light scattering (DLS) system (Brookhaven Instruments BI-2000SM goniometer equipped with a PCI BI-9000AT digital correlator). The laser wavelength is 633 nm. The detector is located at the scattering angle of 90 °. The DLS provides the hydrodynamic size distribution of the particles in the solution by the autocorrelation calculation of the Brownian motion of particles. Sizes distribution of Ag nanoparticles is shown in figure 6. The diameter for most nanoparticles is ~52.1 nm. There are also minor proportions in the size distribution. Especially the nanoparticles less than 10 nm in the solution are common for the synthesis process. In addition, the sizes of particles obtained from DLS are always greater than those from electron microscopes due to hydrodynamic effects in the water [31].

**3.2 Far-field Nanoparticle Observation**

The prepared nanoparticle solutions can be examined using a standard dark field microscopy. Shown in Figure 7(a), the silver nanoparticles were recorded with an exposure time of 3.3 ms under the illumination of white light. The image under the illumination of a single laser pulse (532nm, 10ns duration) was shown in Figure 7(b). The sizes of spots in the images represented the silver nanoparticles is nearly ~ 0.5 μm.

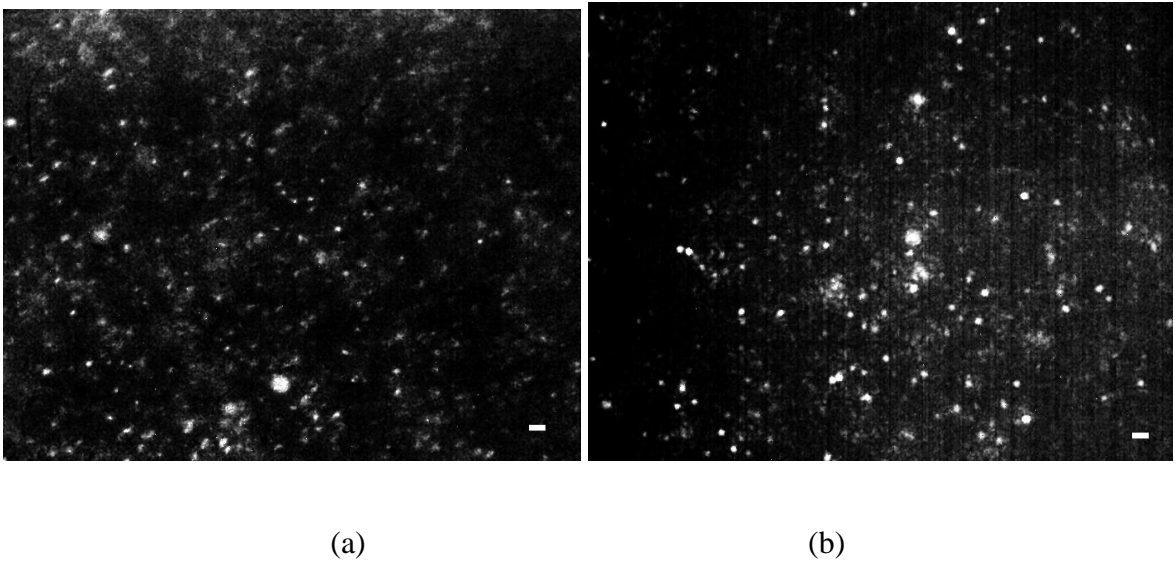

(a) (b)

*Figure 7. Dark field images of Ag nanoparticles illuminated by (a) a white light source. Exposure time for the camera is 4 ms. (b) a laser pulse at 532nm, exposure time equals to the pulse duration of laser, which is about 10ns. The bar is 2 μm.*

Following the analysis used in μPIV, the final sizes of spots recorded by CCD can be estimated by the following equation [15],

$$d_e = \left[ M^2 d_p^2 + 5.95 \cdot (M+1)^2 \lambda^2 \left( \frac{1}{2NA} \right)^2 \right]^{1/2} \quad (3)$$

where $d_e$ is the final imaged particle diameter, M is image magnification (100 ×), $d_p$ is the particle diameter (~35 nm), λ is the light wavelength (500 nm or 532 nm), , NA is the numerical aperture of objective lens (1.4). So the final imaged particle diameter of silver nanoparticles is ~0.55pixles in the images, which corresponding to 0.55 μm in the images.

## 3.3 Velocity Measurement

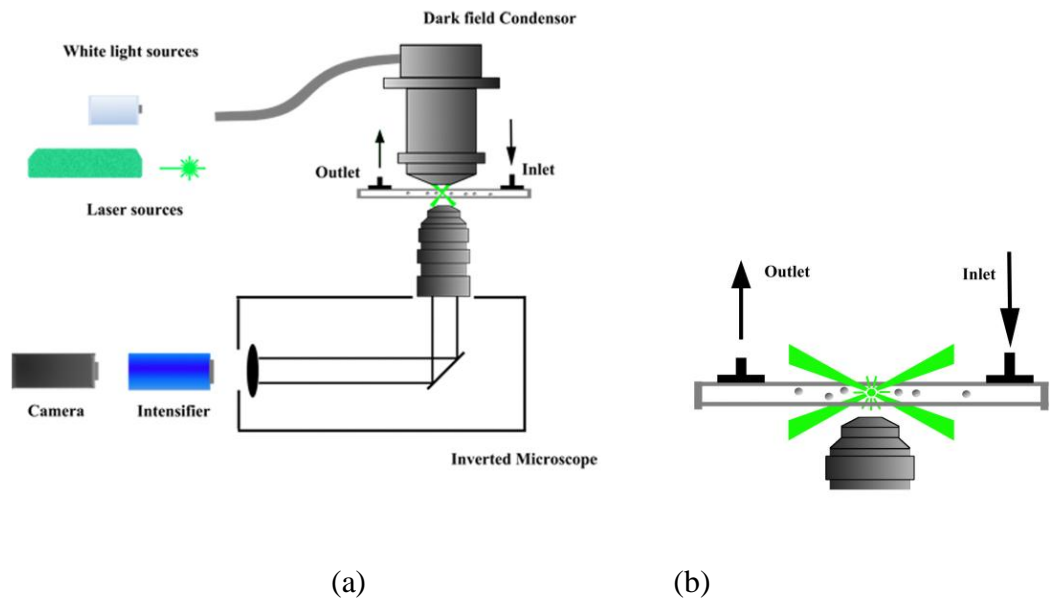

(a)            (b)

*Figure 8. (a) The diagram of nano-PIV setup used dark field illumination, (b) the schematics of the micro channel.*

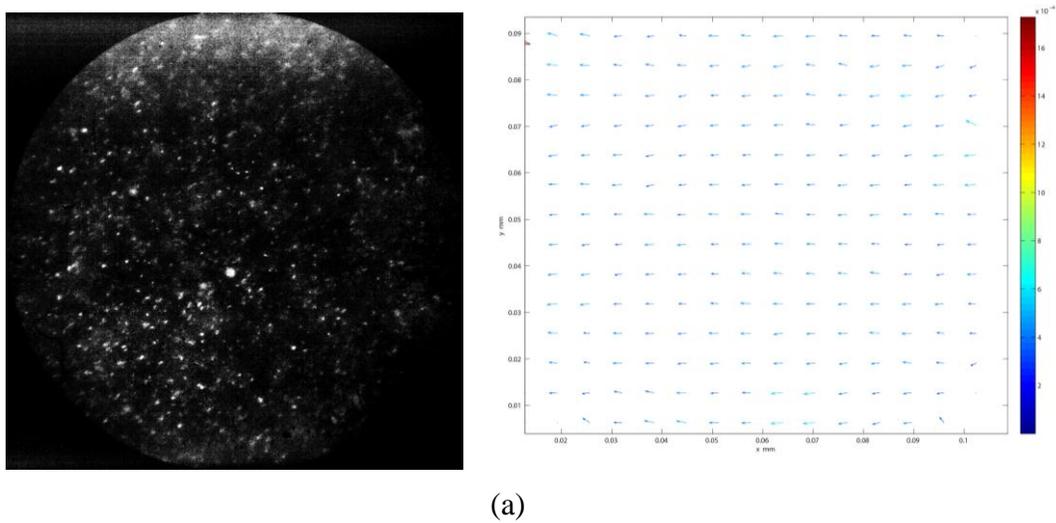

(a)

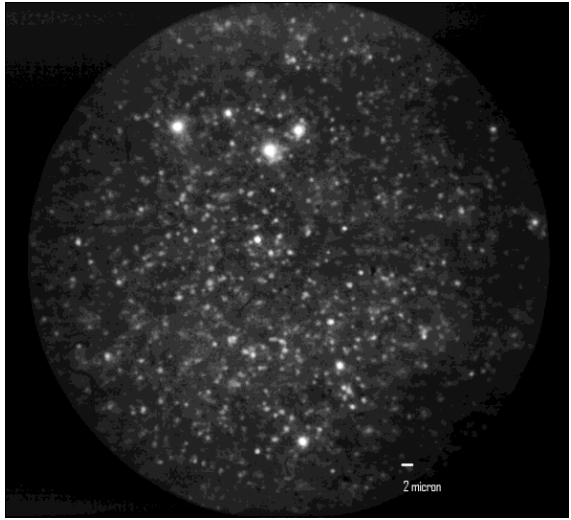 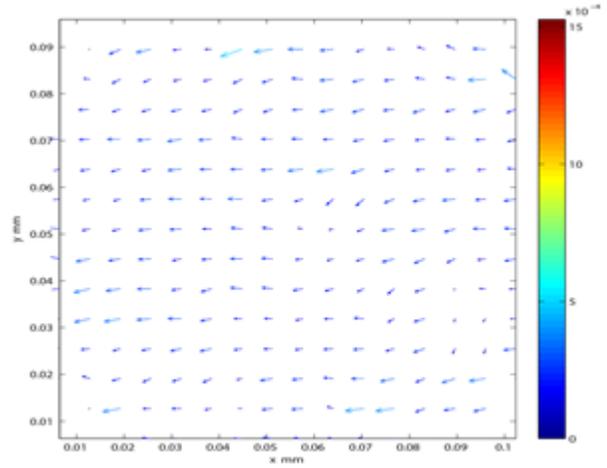

*(b)*

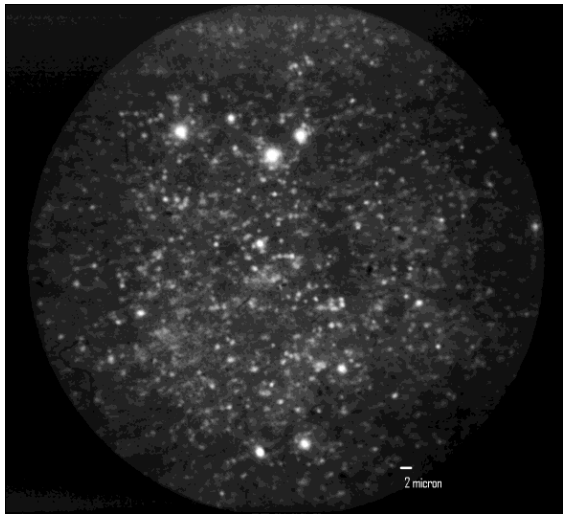 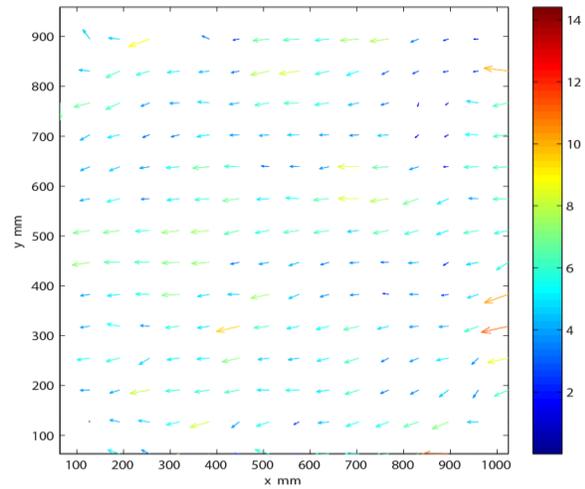

*(c)*

*Figure 9. Experimental measurement of nanoparticle movement in a micro channel using white light dark- field illumination. (a) Image of nanoparticles at an exposure of 2 milliseconds and Instantaneous velocity map of the flow (in m/s). Flow speed is 0.01ml/sec, (b) Flow speed is 0.08ml/sec, (c) flow speed is 0.16ml/sec.*

A far field plasmonic enhanced nano-Particle Image Velocimetry (nPIV) has been demonstrated by integrating the common PIV technique with the single nanoparticle tracking technique. As shown in Figure 8, a laser or white light source has been focused on a micro channel (shown in Figure 8 B) by a dark field condenser. The flow seeded with nanoparticles has been pumped through the channel using syringe pump. A fast camera (TSI- HS650) with an image intensifier has been used to capture scattering light from the nanoparticles. The movement of the flow is accompanied by the displacement of the seeding particles. Since only the scattering light from the nanoparticles can be detected by the camera, it is straightforward to locate those nanoparticles. By comparing two consecutive images of the flow field with auto-/cross-correlation technique, we can follow the movement of the nanoparticles so that a 2D velocity profile has been obtained as shown in Figure 9. The flow velocity is about 0.5 mm/s.

## 4. Conclusions

In this paper, a novel far-field plasmonic resonance enhanced nanoparticle-seeded Particle Image Velocimetry (nPIV) has been demonstrated to measure the velocity in a micro channel. Calculations based on Discrete Dipole Approximation are conducted to optimize light scattering from chemically synthesized silver nanoparticles. Optimum scattering wavelength is specified for the nanoparticles in two media: water and air. The diffraction-limited plasmonic resonance enhanced images of silver nanoparticles at different diameters have been recorded. By using standard PIV techniques, the velocity profiles within the micro channel have been determined from the images. Velocity of 0.5mm/s has been successfully measured by the technique.

# 5. Acknowledgements

The work is supported by the University of Tennessee and Joint Directed Research Direction (JDRD) of the Science Alliance.